\documentclass[12pt]{article}
\usepackage{a4wide}
\usepackage{amssymb}
\usepackage{amsmath}
\usepackage{graphicx}
\usepackage{mathdots}
\usepackage{slashed}
\usepackage{verbatim}
\usepackage{cite}
\usepackage{xcolor}
\usepackage[utf8]{inputenc}

\usepackage{hyperref}
\usepackage{cleveref}

\usepackage{IEEEtrantools}

\def\makeatletter{\catcode`\@=11}% 11:letter
\makeatletter
\def\mathbox#1{\hbox{$\m@th#1$}}%
\def\math@ccstyles#1#2#3#4#5#6#7{{\leavevmode
      \setbox0\mathbox{#6#7}%
      \setbox2\mathbox{#4#5}%
      \dimen@ #3%
      \baselineskip\z@\lineskiplimit#1\lineskip\z@
      \vbox{\ialign{##\crcr
             \hfil \kern #2\box2 \hfil\crcr
             \noalign{\kern\dimen@}%
             \hfil\box0\hfil\crcr}}}}
\def\mathaccstyles{\math@ccstyles\maxdimen}
\def\maththroughstyles{\math@ccstyles{-\maxdimen}}
\def\unity%
 {\maththroughstyles{.45\ht0}\z@\displaystyle {\mathchar"006C}\displaystyle 1}
 
\usepackage{tikz}
\usetikzlibrary{positioning,arrows}
\usetikzlibrary{decorations.pathmorphing,calc}
\usetikzlibrary{decorations.markings}
\newif\ifmirrorsemicircle
\usepackage{float}
\usepackage{ifthen}
\usepackage{enumerate}
\usepackage{amsmath}
\usepackage{amssymb}
\usepackage[mathscr]{eucal}
\usepackage[errorshow]{tracefnt}
\usepackage{amsmath}
\usepackage{slashed}
\usepackage{amssymb}
\usepackage{array}
\usepackage{amsthm}
\usepackage{eucal}
\usepackage{epsf}
\usepackage{epsfig}
\usepackage{psfrag}
\usepackage{multirow}
\usepackage{wasysym} 
\usepackage{tikz-cd} 
\usepgfmodule{shapes}
\usetikzlibrary{backgrounds, arrows,calc,shapes,decorations.pathreplacing, decorations.markings, automata,positioning}
\usepackage{verbatim}
 \newcommand{\dirac}[1]{\langle #1 \rangle}

 \numberwithin{equation}{section}
\begin{document}

\begin{flushright}\footnotesize

\texttt{}
\vspace{0.6cm}
\end{flushright}

\mbox{}
\vspace{0truecm}
\linespread{1.1}

%%%%%%%%%%%%%%%%%
\centerline{\LARGE \bf Continuous symmetry defects}
\medskip

\centerline{\LARGE \bf  and brane/anti-brane systems }
\medskip

%\centerline{\LARGE \bf } 

\vspace{.5cm}

 \centerline{\LARGE \bf }

\vspace{1.5truecm}

\centerline{
    {\bf Hugo Calvo${}^{a,b}$} \footnote{calvohugo@uniovi.es},
    { \bf Francesco Mignosa ${}^{a,b}$} \footnote{francesco.mignosa02@gmail.com}
        {\bf and}
    { \bf Diego Rodriguez-Gomez${}^{a,b}$} \footnote{d.rodriguez.gomez@uniovi.es}}

\vspace{1cm}
\centerline{{\it ${}^a$ Department of Physics, Universidad de Oviedo}} \centerline{{\it C/ Federico Garc\'ia Lorca  18, 33007  Oviedo, Spain}}
\medskip
\centerline{{\it ${}^b$  Instituto Universitario de Ciencias y Tecnolog\'ias Espaciales de Asturias (ICTEA)}}\centerline{{\it C/~de la Independencia 13, 33004 Oviedo, Spain.}}
\vspace{1cm}

\centerline{\bf ABSTRACT}
\medskip

We explicitly compute correlation functions with the insertion of a continuous symmetry defect in bosonic field theories. To recover the expected action, the definition of the defect must be modified to include a specific contact term. This can be regarded as a singular background gauge field for the global symmetry. It can be traced to the definition of the generating functional for current correlators, where the source is akin to a background gauge field for the symmetry. For holographic theories, it has been proposed that continuous symmetry defects are realized in terms of non-BPS $D(q-1)$ branes. We argue that these can be regarded as $Dq/\overline{Dq}$ system and show its application to the case of the baryonic symmetry in the Klebanov-Witten theory. The  $Dq/\overline{Dq}$ can be regarded as a particular regularization of the defect, holographically realizing the field theory discussion.

\noindent

\newpage

\tableofcontents

\section{Introduction}

Symmetries play a prominent role in Quantum Field Theory (QFT). They impose selection rules constraining physical processes, while their anomalies and possible spontaneous symmetry breaking patterns allow to constraint possible endpoints of  renormalization group flows. It is then of great importance to exploit all possible symmetries that a theory possesses. Over the last decade, starting with \cite{Gaiotto:2014kfa}, it has been realized that symmetries can be associated to the existence of a sector of defect operators supported on submanifolds of spacetime on which they depend only topologically. This not only includes the familiar case of continuous global symmetries and their associated conserved currents through Noether's theorem, but naturally includes discrete symmetries as well. Moreover, it is natural to generalize the notion of symmetry to operators supported on submanifolds of arbitrary dimension --leading to the so-called higher-form symmetries-- and to cases involving more exotic fusion rules than those dictated by a group structure --leading to the so-called non-invertible symmetries.

Stemming from the original work in \cite{Gaiotto:2014kfa}, over the last decade there has been great progress in the study of generalized symmetries. These were studied in the context of category theory \cite{Bartsch:2022mpm, Bartsch:2022ytj, Bhardwaj:2022yxj, Bartsch:2023pzl, Bartsch:2023wvv, Bhardwaj:2023wzd, Bhardwaj:2023ayw, Bullimore:2024khm}, as well as employing the framework of the symTFT \cite{Freed:2012bs,Ji:2019eqo, Gaiotto:2020iye, Apruzzi:2021nmk, Freed:2022qnc}, which allows to separate the global symmetry structure from the field theory dynamics. Generalized global symmetries were also explored in the context of holography \cite{Apruzzi:2021nmk, Bergman:2020ifi, Bergman:2022otk, Apruzzi:2022rei,Heckman:2022xgu, GarciaEtxebarria:2022vzq, Bah:2023ymy, Argurio:2024oym, Waddleton:2024iiv, Bergman:2024its}, where the role of the symTFT is played by the topological sector of the reduction of the supergravity theory over the dual holographic background. Its topological operators are described by branes, which are dual to charged or topological operators of the dual theory depending on the boundary conditions of the supergravity fields at the boundary of $AdS$. Very recently, it was proposed in \cite{Bergman:2024aly} that the holographic realization of (at least a class of continuous) symmetry operators is in terms of non-BPS $D$-branes\footnote{See \cite{Gutperle:2001mb, Cvetic:2023plv} for an alternative proposal given by flux branes.}, which naturally link with the holographic dual of charged operators. These branes are unstable and their dynamics is described by a tachyon effective field theory. As argued in \cite{Bergman:2024aly}, the presence of background RR field strengths introduces a hidden parameter in their worldvolume, so that in the tachyon vacuum, the non-BPS brane reduces to a topological operator measuring a U(1) charge.

The topological operators associated to a certain symmetry can be effectively described as $e^{i\alpha Q[M]}$, where $\alpha$ is a parameter labelling the operator\footnote{This takes values in a suitable domain, for instance $\mathbb{R}/\mathbb{Z}$ for $U(1)$ symmetries or $\mathbb{Z}_N$ for $\mathbb{Z}_N$ symmetries.} and $Q[M]$ counts the number of charges inside the surface $M$. In the vanilla case of continuous $p$-form symmetries in $d$-dimensional QFT's, this admits a more explicit description. In that case, the familiar Noether theorem ensures the existence of a $p+1$-form current $j_{p+1}$ satisfying the conservation equation $d\star j_{p+1}=0$. This automatically allows to explicitly construct the associated symmetry operators as $e^{i\alpha\int_{M_{d-p-1}}\star j}$. By virtue of the conservation equation, this operator does not depend on $M_{d-p-1}$ as long as it does not cross charged operators. This is however a classical operator which may require adjustments at the quantum level. This was first addressed only very recently in \cite{Bah:2024ucp}. Motivated by this, we set out to explicitly computing correlation functions in the presence of defects for the case of $0$-form symmetries for bosonic theories. As argued in \cite{Bah:2024ucp}, in those cases we find that a contact term is needed to recover the correct action of the defect. Actually, this contact term can be traced back to the fact that the relevant generating functional for current correlators is akin to adding a background field for the global symmetry. Since for bosonic theories the kinetic terms are quadratic, this automatically generates the relevant contact terms. 

As usual in QFT, computations require some form of regularization. Although it is natural to use a cut-off in momentum space, as argued in \cite{Bah:2024ucp}, another convenient regulator is to thicken the defect. For holographic theories, where the holographic description of symmetry operators is in terms of non-BPS $D$-branes, this resonates with the well-known fact that non-BPS $D(q-1)$ branes can be also seen as the endpoint of a decay process involving the tachyon dynamics of a $Dq/\overline{Dq}$ system \cite{Sen:1998ii, Sen:1998tt, Alishahiha:2000du}. The tachyon is in this case complex and fluctuations of a kink solution of its real part in flat space was found to give rise to a non-BPS $D(q-1)$ brane theory localized at the position of the center of kink. Motivated by this we study the dynamics of $Dq/\overline{Dq}$ in $AdS$. While the worldvolume dynamics cannot be explicitly solved, we argue that the endpoint of the system is a $U$ shaped $Dq$ brane hanging from the boundary. It is natural to identify this system in the limit in which the $U$ hangs very little from the boundary with the non-BPS $D(q-1)$ brane. In order to make this connection more explicit, we consider in detail the case of the baryonic symmetry in the Klebanov-Witten theory, whose gravity dual background is in terms of $AdS_5\times T^{1,1}$. The baryon corresponds to a $D3$ brane wrapping the 3-cycle of the $T^{1,1}$ and extended along the radial direction of $AdS$, and naturally links with the symmetry operator described by non-BPS $D4$ brane wrapping a 3-manifold at the boundary and the 2-cycle of the $T^{1,1}$. Upon regarding the non-BPS $D4$ as a $D5/\overline{D5}$, one can make explicit the connection to the baryonic symmetry. Indeed, the $D5$ in isolation is well-known to represent a domain wall increasing the relative rank of the gauge group \cite{Gubser:1998fp}. As a consequence, the $U$-shaped $D5$ brane is only sensitive to baryon operators, as expected for a baryon symmetry operator, which leave behind a string when dragged across the defect. Moreover, a similar argument to that in \cite{Green:1996dd} shows that a gauge transformation of the worldvolume gauge field on the $D5$ induces a shift of the baryonic gauge field. As a consequence, we can regard the effect of the crossing of the baryon $D3$ through the defect as a background for the baryon symmetry gauge field in parallel to the QFT discussion.

The outline of this note is as follows. In section \ref{divergences}, we study correlation functions in the presence of continuous symmetry operators in (free) field theory for $U(1)$ and shift symmetries. We find that in order to reproduce the correct action of the symmetry defect, its definition must involve a particular contact term whose effect is to complete derivatives into covariant derivatives with a singular background gauge field. We argue that this traces back to the fact that the relevant generating functional for current correlators is akin to introducing a background gauge field for the symmetry. While for fermionic theories with non-derivative couplings this is clear, for bosons --with two-derivative kinetic terms-- this has the effect of automatically producing the expected contact terms. In section \ref{holography} we turn to holographic symmetry defects. We argue that $Dq/\overline{Dq}$ in $AdS$ decays into a $U$-shaped $Dq$ which, in the limit of small size, we identify with the non-BPS $D(q-1)$. We then particularize to the baryonic symmetry in Klebanov-Witten and argue for the precise connection to the baryonic symmetry. We end in section \ref{conclusions} with a summary and outlook. Finally, we leave several technical details to the appendices \ref{Formulae}, \ref{Calculationsofintegrals} and \ref{appU}.

\section{Continuous symmetry defects in field theory}\label{divergences}

We consider euclidean field theories of a single scalar field in $d$ dimensions living on $\mathbb{R}^d$ parametrized by the coordinates  $\{x^0,\cdots,x^{d-1}\}$ which enjoy either a shift $\mathbb{R}$ 0-form symmetry or a $U(1)$ global 0-form symmetry. The actions are respectively

\begin{equation}
S_{\rm shift}=\int d^dx\, \frac{1}{2}\partial\phi^2+V(\partial\phi)\,;\qquad S_{U(1)}=\int d^dx\, |\partial\phi|^2+V(|\phi|)\,.
\end{equation}
In the following we will mostly consider the free case by setting $V=0$.

$S_{\rm shift}$ is invariant under the continuous $\mathbb{R}$ symmetry $\phi\rightarrow \phi+\alpha$. In turn, $S_{U(1)}$ is invariant under the continuous $U(1)$ symmetry $\phi\rightarrow e^{i\alpha}\phi$. Noether's theorem ensures the existence of a conserved 1-form current $j=j_{\mu}dx^{\mu}$. In each case, it reads

\begin{equation}
j_{\mu}^{\rm shift}=-i\partial_{\mu}\phi\,,\qquad j_{\mu}^{U(1)}=\phi\partial_{\mu}\phi^{\star}-\phi^{\star}\partial_{\mu}\phi\,.
\end{equation}

In modern language the symmetry is implemented by co-dimension 1 defects supported on a $d-1$ dimensional manifold $M_{d-1}$ which depend only topologically on $M_{d-1}$. Naively, the symmetry operators are

\begin{equation}
U_{\alpha}(M_{d-1})=e^{i\alpha\int_{M_{d-1}} \star j}\,,
\end{equation}
where $\alpha$ labels the symmetry defect. For concreteness, we will consider planar defects located at $x^0=0$, for which $M_{d-1}=\mathbb{R}^{d-1}$ parametrized by $\vec{x}=(x^1,\cdots,x^{d-1})$. Then, the naive symmetry defect can be written as 

\begin{equation}
\label{defectnaive}
U_{\alpha}=e^{i\alpha\int d^{d-1}\vec{x} j_0}=e^{i\alpha\int d^dx\,\delta(x^0)\, j_0}\,.
\end{equation}

We will be interested on computing correlation functions with the insertion of the defect $\langle \mathcal{F}(\phi,\phi^{\star})\,U_{\alpha}\rangle$,

where $\mathcal{F}$ is some (generically non-local) composite in the field and its conjugate. We will organize the  computation by expanding the exponential in powers of $\alpha$ as

\begin{equation}
\label{correlator}
\langle \mathcal{F}\,U_{\alpha}\rangle=\langle \mathcal{F}\rangle+i\alpha \int d^dz_1\,\delta(z_1^0)\,\langle \mathcal{F}\,j^0(z_1)\rangle-\frac{\alpha^2}{2}\int d^dz_1\int d^dz_2\,\delta(z_1^0)\delta(z_2^0)\,\langle \mathcal{F}\,j^0(z_1)\,j^0(z_2)\rangle+\cdots\,.
\end{equation}

\subsection{$U(1)$ symmetry}

Let us consider the two point function in the presence of the symmetry defect, which corresponds to $\mathcal{F}=\phi(x)\,\phi^{\star}(y)$.

The $\mathcal{O}(\alpha)$ contribution to \eqref{correlator} comes from the 3-point correlator $\langle \phi(x)\phi^{\star}(y)\,j^{\mu}(z)\rangle$. Since the theory is free, we can just factorize the correlation function as 
\begin{equation}
\dirac{\phi(x)\phi^\ast (y) j^\mu(z)}= \partial^\mu_z\dirac{\phi(x) \phi^\star(z) }\dirac{\phi^\star(y) \phi(z)}-  \dirac{\phi(x) \phi^\star(z) }\partial^\mu_z\dirac{\phi^{\star} (y) \phi(z)}\,.
\end{equation}
This can be explicitly written in terms of free field 2-point functions. The result is (see appendix \ref{Formulae}) 

\begin{align}\label{Integral1}
\dirac{\phi(x)\phi^\star (y) j^\mu(z)}= (d-2)C_d^2\left(\frac{(x-z)^\mu}{|x-z|^d|y-z|^{d-2}}-\frac{(y-z)^\mu}{|x-z|^{d-2}|y-z|^d}\right)\,.
\end{align}
Diagramatically this corresponds to the diagram in fig. \ref{phibarphi-current} below.
\begin{figure}[h!]
\centering
\includegraphics[scale=.15, trim={0.2cm, 7cm, 4cm, 7cm}, clip]{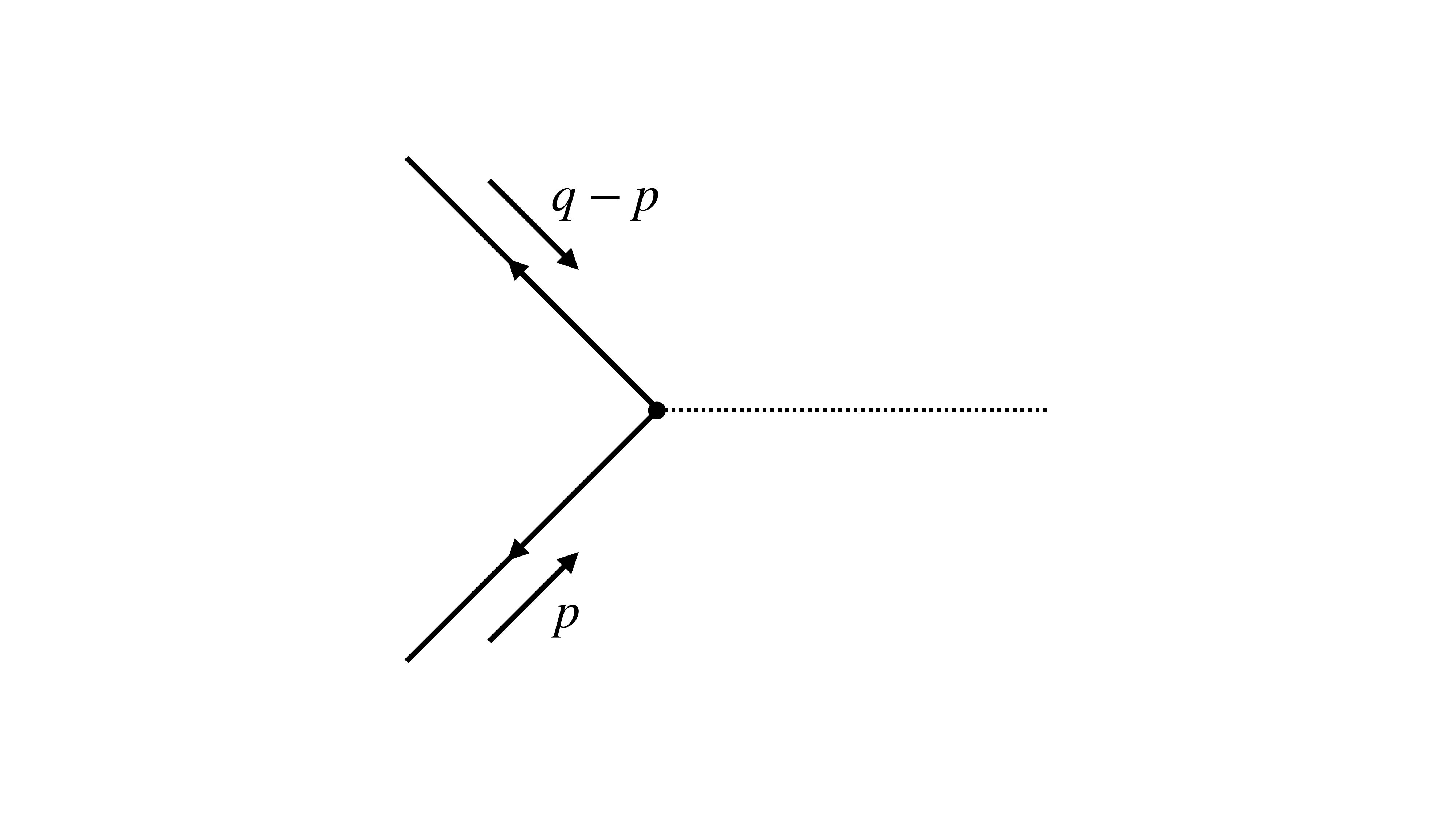}
\caption{Tree level diagrams for $\mathcal{O}(\alpha)$ contribution.}
\label{phibarphi-current}
\end{figure}
Inserting this into eq. \eqref{correlator}, the $\mathcal{O}(\alpha)$ contribution reads (see appendix \ref{Calculationsofintegrals} for details)

\begin{equation}
\int d^dz\,\delta(z^0)\,\langle \phi(x)\phi^{\star}(y)\,j^0(z)\rangle=\frac{C_d}{|x-y|^{d-2}}\,\frac{1}{2}\,\big({\rm sign}(x^0)-{\rm sign}(y^0)\big)\,.
\end{equation}
This can be neatly re-written as

\begin{equation}
\int d^dz\,\delta(z^0)\,\langle \phi(x)\phi^{\star}(y)\,j^0(z)\rangle=\frac{1}{2}\,\big({\rm sign}(x^0)-{\rm sign}(y^0)\big)\langle \phi(x)\,\phi^{\star}(y)\rangle\,.
\end{equation}

We now turn to the $\mathcal{O}(\alpha^2)$ contribution to eq. \eqref{correlator}, which requires to compute $\dirac{\phi(x)\phi^\star(y) j^\mu(z_1) j^\nu (z_2)}$.  Diagramatically, since that we are interested in connected diagrams (\textit{i.e.} not those corresponding to Wick contractions for $\langle \phi(x)\phi(y)\rangle\langle j^{\mu}(z_1)\,j^{\nu}(z_2)\rangle$), the relevant ones are those in fig. \ref{phibarphi-currentcurrent} below.

\begin{figure}[h!]
\centering
\includegraphics[scale=.17, trim={0.2cm, 6cm, 4cm, 6cm}, clip]{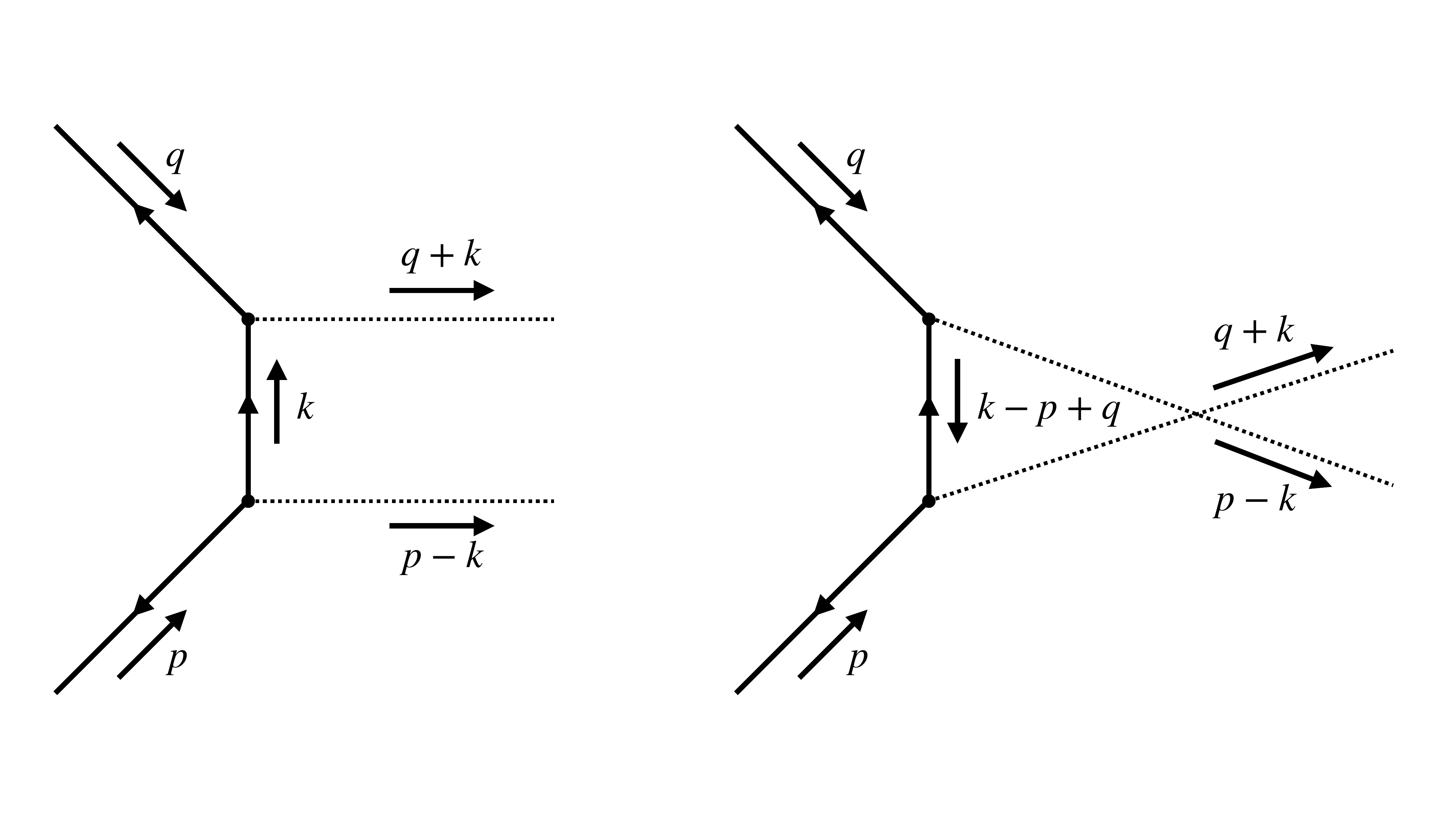}
\caption{Tree level diagrams for $\mathcal{O}(\alpha^2)$ contribution.}
\label{phibarphi-currentcurrent}
\end{figure}
We can easily compute the correlator using Wick contractions (and focusing on those connected). The result is 

\begin{equation}
\dirac{\phi(x)\phi^\star(y) j^\mu(z_1) j^\nu (z_2)}=\mathcal{A}^{\mu\nu}+\mathcal{B}^{\mu\nu}\,,
\end{equation}
where $\mathcal{A}_{\mu\nu},\,\mathcal{B}_{\mu\nu}$ correspond to the diagrams above and read explicitly

\begin{eqnarray}
\nonumber \mathcal{A}_{\mu\nu}&=&   (d-2)^2C_d^3\, \frac{(x^{\mu}-z_1^{\mu})\,(z_1^{\nu}-z_2^{\nu})}{|z_2-y|^{d-2}\,|x-z_1|^d\,|z_1-z_2|^d}   + (d-2)^2C_d^3 \,\frac{(x^{\mu}-z_1^{\mu})\,(z_2^{\nu}-y^{\nu})}{|z_1-z_2|^{d-2}\,|x-z_1|^d\,|z_2-y|^d} \\ \nonumber && -(d-2)C_d^3\,\frac{1}{|z_2-y|^{d-2}\,|x-z_1|^{d-2}\,|z_1-z_2|^d}\, \Big(\delta^{\mu\nu}-d\frac{(z_1^{\mu}-z_2^{\mu})\,(z_1^{\nu}-z_2^{\nu})}{|z_1-z_2|^2}\Big)
\\ &&  +(d-2)^2C_d^3  \frac{(z_1^{\mu}-z_2^{\mu})\, (z_2^{\nu}-y^{\nu})}{|x-z_1|^{d-2}\,|z_2-y|^d\,|z_1-z_2|^d}\,,
\end{eqnarray}
and

\begin{eqnarray}
\nonumber \mathcal{B}_{\mu\nu}&=&   -(d-2)^2C_d^3\, \frac{(x^{\nu}-z_2^{\nu})\,(z_1^{\mu}-z_2^{\mu})}{|z_1-y|^{d-2}\,|x-z_2|^d\,|z_1-z_2|^d}   + (d-2)^2C_d^3 \,\frac{(x^{\nu}-z_2^{\nu})\,(z_1^{\mu}-y^{\mu})}{|z_1-z_2|^{d-2}\,|x-z_2|^d\,|z_1-y|^d} \\ \nonumber &&- (d-2)C_d^3\,\frac{1}{|z_1-y|^{d-2}\,|x-z_2|^{d-2}\,|z_1-z_2|^d}\, \Big(\delta^{\mu\nu}-d\frac{(z_1^{\mu}-z_2^{\mu})\,(z_1^{\nu}-z_2^{\nu})}{|z_1-z_2|^2}\Big)
\\  && -(d-2)^2C_d^3  \frac{(z_1^{\nu}-z_2^{\nu})\, (z_1^{\mu}-y^{\mu})}{|x-z_2|^{d-2}\,|z_1-y|^d\,|z_1-z_2|^d}\,.
\end{eqnarray}

For our purposes we need the integrated $\mu=\nu=0$ component with $z_1^0=z_2^0=0$. This is
\begin{align}
\label{contributionD1}
&-2(d-2)C_d^3\int d^{d-1}\vec{z}_1\int d^{d-1}\vec{z}_2\Big\{ (d-2) \,\frac{x^0\,y^0}{|z_1-z_2|^{d-2}\,|x-z_1|^d\,|z_2-y|^d}+\\
&+\,\frac{1}{|z_2-y|^{d-2}\,|x-z_1|^{d-2}\,|z_1-z_2|^d}\Big\}\,.\nonumber%+\frac{1}{|x-y|^{d-2} |z_1-z_2|^{2d-2}}\Big\}\,.\nonumber
\end{align}
The full integral has a short distance singularity $z_1\rightarrow z_2$. Upon closer inspection, one sees that only the second integral diverges. To handle the divergence, let us re-write

\begin{equation}
\frac{C_d(d-2)}{|z_1-z_2|^d}= \delta(z_1^0)\delta(z_2^0) \frac{C_d(d-2)}{|z_1-z_2|^d}\left(\delta_{00}-d\frac{(z_1-z_2)_0(z_1-z_2)_0}{|z_1-z_2|^2}\right)=\delta(z_1^0)\delta(z_2^0) \int \frac{d^dk}{(2\pi)^d} e^{ik\cdot(z_1-z_2)}\frac{k_0^2}{k^2}\,.
\end{equation}
Hence the integral becomes

\begin{align*}
-2 \int d^{d-1}z_1d^{d-1}z_2\int \frac{d^dp}{(2\pi)^d}\frac{d^dq}{(2\pi)^d}\frac{d^dk}{(2\pi)^d}e^{ip\cdot(x-z_1)}e^{iq\cdot(y-z_2)}e^{ik(z_1-z_2)}\frac{k_0^2}{k^2p^2 q^2}\,.
\end{align*}
Integrating over $z_1, z_2$

\begin{equation}
-2\int \frac{dp^0}{2\pi} \frac{dq^0}{2\pi} \frac{d^d k}{(2\pi)^d}e^{ip^0 x^0+iq^0y^0}e^{i\vec{k}\cdot (\vec{x}- \vec{y})}\frac{k_0^2}{((p_0)^2+\vec{k}^2)((q_0)^2+\vec{k}^2)k^2}\,.
\end{equation}
Integrating over $p_0, q_0$ we find

\begin{equation}
-\int \frac{d^{d-1}k}{(2\pi)^{d-1}}\int \frac{dk^0}{2\pi} e^{i\vec{k}\cdot (\vec{x}- \vec{y})} e^{-|\vec{k}|(|x^0|+|y^0|)}\frac{k_0^2}{2\vec{k}^2 ((k^0)^2+\vec{k}^2)}\,.
\end{equation}
The integration over $k_0$ shows the aforementioned divergence. We can separate the divergent part from the finite part as

\begin{equation}
\int_{-\infty}^{\infty} \frac{dk_0}{2\pi} \frac{k_0^2}{k_0^2+\vec{k}^2}=\lim_{\Lambda\rightarrow\infty} \int_{-\Lambda}^{\Lambda} \frac{dk_0}{2\pi} \left(1-\frac{\vec{k}^2}{k_0^2+\vec{k}^2}\right)= \frac{\Lambda}{\pi}-\frac{\vec{k}^2}{2|\vec{k}|}\,,
\end{equation}
where we introduced a regularization for the $\delta(0)$ function

\begin{equation}
\label{regularizeddelta}
\delta(0)= \int_{-\infty}^\infty \frac{dk^0}{2\pi} = \lim_{\Lambda\rightarrow\infty}\int_{-\Lambda}^\Lambda\frac{dk^0}{2\pi}=\lim_{\Lambda\rightarrow \infty}\frac{\Lambda}{\pi}\,.
\end{equation}
Thus, the integral becomes

\begin{equation}
-\delta(0)\, \int \frac{d^{d-1}k}{(2\pi)^{d-1}} \frac{e^{-|\vec{k}|(|x^0|+|y^0|)+i\vec{k}\cdot (\vec{x}- \vec{y})}}{2\vec{k}^2} +\frac{1}{2} \int \frac{d^{d-1}k}{(2\pi)^{d-1}} \frac{e^{-|\vec{k}|(|x^0|+|y^0|)+i\vec{k}\cdot (\vec{x}- \vec{y})}}{2|\vec{k}|}\,.
\end{equation}

Let us now turn to the first integral. In momentum space

\begin{equation}
2\int d^{d-1}z_1 d^{d-1}z_2 \int \frac{d^dp}{(2\pi)^d} \frac{d^dq}{(2\pi)^d}\frac{d^dk}{(2\pi)^d} e^{i p\cdot (x-z_1)}e^{i q\cdot (y-z_2)}e^{i k\cdot (z_1-z_2)}\frac{p^0q^0}{p^2q^2k^2}\,.
\end{equation}
The two integrals over $z_1, \,z_2$ give $\delta(\vec{p}-\vec{k})\delta(\vec{q}+\vec{k})$. Hence

\begin{equation}
2 \int \frac{dp^0}{2\pi} \frac{dq^0}{2\pi}\frac{dk^0}{2\pi} \frac{d^{d-1}\vec{k}}{(2\pi)^{d-1}} e^{i p^0 x^0+i q^0 y^0}e^{i\vec{k}\cdot (\vec{x}- \vec{y})}\frac{p^0q^0}{((p^0)^2+\vec{k}^2)((q^0)^2+\vec{k}^2)((k^0)^2+\vec{k}^2)}\,.
\end{equation}
Integrating sequentially on $p^0$, $q^0$ and $k^0$, one finds

\begin{equation}
%-\frac{\text{sgn}(x^0y^0)}{2}\int \frac{d^dk}{(2\pi)^{d-1}}\frac{e^{-|\vec{k}||x^0|-|\vec{k}||y^0|+i \vec{k}\cdot\vec{x}-i \vec{k}\cdot\vec{y}}}{k^2}=
-\frac{\text{sgn}(x^0y^0)}{2} \int \frac{d^{d-1}k}{(2\pi)^d} \frac{e^{-|\vec{k}||(x^0|+|y^0|)+i \vec{k}\cdot(\vec{x}-\vec{y})}}{2|\vec{k}|}\, .
\end{equation}

Finally, summing together the finite parts and using the formulae in Appendix \ref{Formulae}, we obtain

\begin{equation}
\Big(\frac{\text{sgn}(x^0)-\text{sign}(y^0)}{2}\Big)^2\, \frac{C_d}{|x-y|^{d-2}}\,.
\end{equation}

As for the divergent part, using that

\begin{equation}
\int d^{d-1} z \dirac{\phi(x)\phi^\star (z)}\dirac{\phi(z)\phi^\star (y)}=
\int \frac{d^{d-1}k}{(2\pi)^{d-1}} e^{i\vec{k}\cdot(\vec{x}-\vec{y})}e^{-|\vec{k}|(|x^0|+|y^0|)} \frac{1}{4 \vec{k}^2}\,,
\end{equation}

it can be rewritten as

\begin{equation}
- 2\delta(0)\,\int d^dz\, \delta(z^0)\,\langle \phi(x)\phi^{\star}(y)\,|\phi|^2(z)\rangle\,.
\end{equation}

Thus we see that the $\mathcal{O}(\alpha)^2$ contribution to eq. \eqref{correlator} would contain

\begin{equation}
\langle \phi(x)\phi^{\star}(y)U_{\alpha}\rangle = \cdots +\alpha^2\,\delta(0)\,\int d^dz\, \delta(z^0)\,\langle \phi(x)\phi^{\star}(y)\,|\phi|^2(z)\rangle+\cdots
\end{equation}
Because of this term, we do not recover the expected action of the symmetry operator. However, the offending term can be cancelled by slightly modifying the defect definition 

\begin{equation}
\widehat{U}_{\alpha}=U_{\alpha}\,e^{-\int d^dx\,J_0^2\,|\phi|^2}\,,
\end{equation}
where $J_0=\alpha\,\delta(x^0)$. Note that the path integral with $\widehat{U}_{\alpha}$ insertions becomes then

\begin{equation}
\langle \mathcal{F}\,\widehat{U}_{\alpha}\rangle=\int\,\mathcal{D}\phi\,\mathcal{F}\,e^{-\int |D\phi|^2}\,,\qquad D_{\mu}\phi=\partial_{\mu}\phi-i\,J_{\mu}\phi\,.
\end{equation}
Moreover, the correlation functions in presence of $\widehat{U}_{\alpha}$ become

\begin{eqnarray}
\langle \phi(x)\phi^{\star}(y)\,\widehat{U}_{\alpha}\rangle&=&\langle \phi(x)\phi^{\star}(y)\rangle+i\alpha \langle \phi(x)\,\phi^{\star}(y)\rangle\,\frac{1}{2}\,\big({\rm sign}(x^0)-{\rm sign}(y^0)\big)\\ \nonumber && +\frac{\alpha^2}{2}\, \langle \phi(x)\phi^{\star}(y)\rangle\,\frac{1}{2}\big({\rm sign}(x^0\,y^0)-1\big)+\cdots \,.
\end{eqnarray}
We recognize here the first terms in the expansion of the expected result

\begin{equation}
\langle \phi(x)\phi^{\star}(y)\widehat{U}_{\alpha}\rangle=\langle \phi(x)\phi^{\star}(y)\rangle\,e^{i\alpha\,(\theta(x^0)-\theta(y^0))}\,.
\end{equation}

\subsubsection{1-loop corrections}

So far we have considered the free theory. Let us now include interactions by turning on a potential (for concreteness $V=\frac{\lambda}{4}|\phi|^4$). Of course, the expectation is that the defect remains topological, and in particular identical to the free theory case. Let us consider just the 1-loop corrections to $\mathcal{O}(\alpha)$, leaving for future work a more exhaustive analysis. To $\mathcal{O}(\alpha)$ we need to consider the corrections to fig. \ref{phibarphi-current}. To 1-loop, the relevant diagrams are in figure \ref{diagrams} below.

\begin{figure}[h!]
\centering
\includegraphics[scale=.17, trim={0.2cm, 7cm, 4cm, 7cm}, clip]{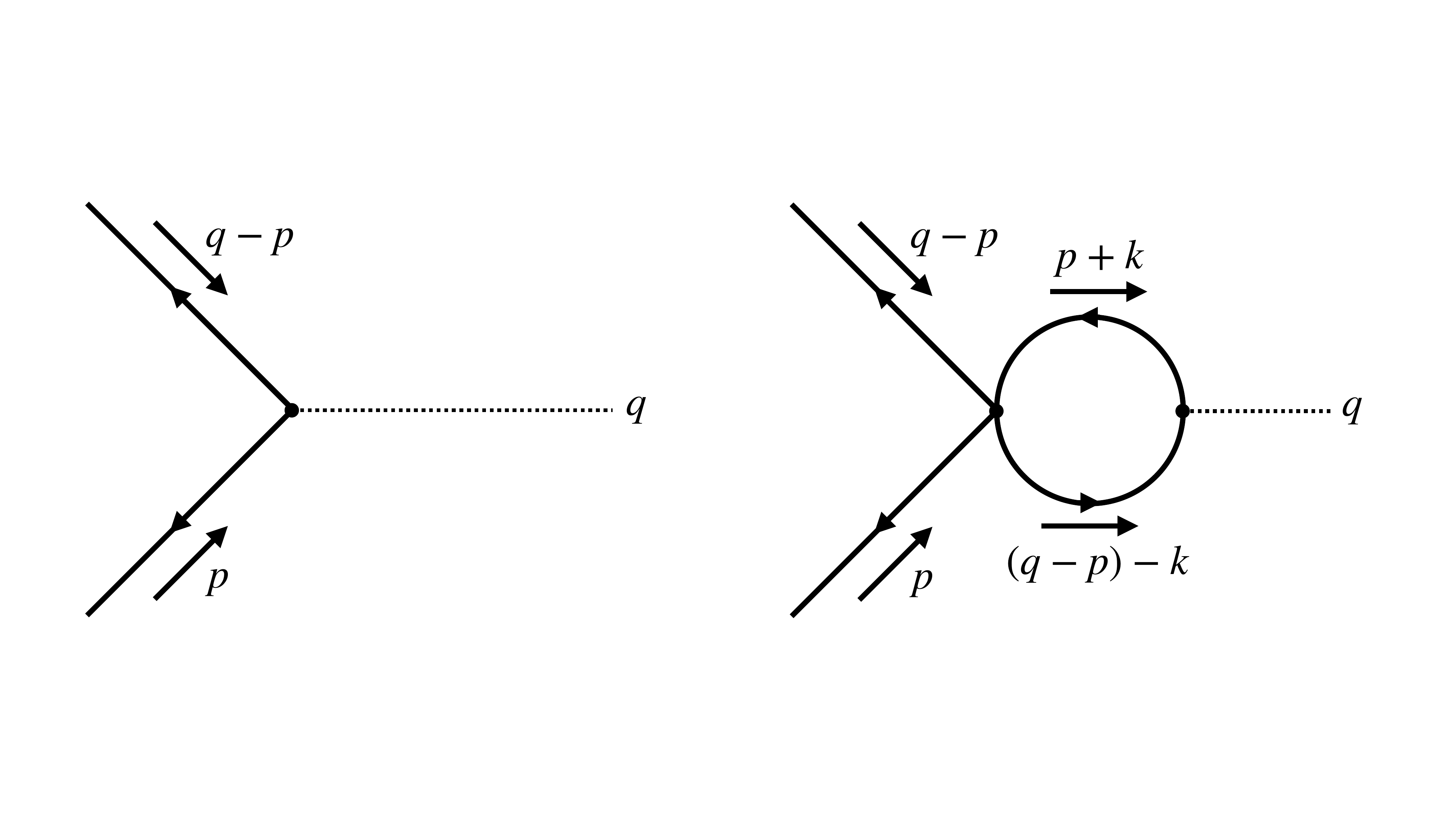}
\caption{One-loop diagrams for $\mathcal{O}(\alpha)$ contribution.}
\label{diagrams}
\end{figure}
The first one is that of the free theory. As for the one-loop correction, in momentum space

\begin{equation}
D_1=-\lambda\,\frac{1}{p^2\,(p-q)^2}\, \int \frac{d^dk}{(2\pi)^d}\, \frac{(2k+2p-q)^{\mu}}{(k+p-q)^2\,(k+p)^2}\,.
\end{equation}
Introducing Feynman parameters and massaging the integral, it reads

\begin{equation}
D_1=-\lambda\,\frac{1}{p^2\,(p-q)^2}\, \int_{-\frac{1}{2}}^{\frac{1}{2}}dy \int \frac{d^d\ell }{(2\pi)^d}\, \frac{2y\,q^{\mu}}{(\ell^2-\Delta)^2}\,,\qquad \Delta=q^2\,\left(y-\frac{1}{2}\right)\,\left(y+\frac{1}{2}\right)\,.
\end{equation}
Since this is linear in $y$, it vanishes by symmetry, so that the result obtained for the free theory extends to one-loop as expected. 

\subsection{Shift symmetry}

Let us now consider the shift symmetry case, and restrict to $d>2$.\footnote{In this regime, the shift symmetry is actually spontaneously broken.} 
Since the shift symmetry does not act linearly on the field it is convenient to consider the vertex operator $V=e^{c\phi(x)}$ for $c$ some constant of appropriate dimension. 

Let us first consider $\langle V(x)\,U_{\alpha}\rangle$. The $\mathcal{O}(\alpha)$ contribution requires to compute $\langle e^{c\phi(x)}\,j^{\mu}(z)\rangle$. Expanding the exponential, it is clear that $\langle e^{c\phi(x)}\,j^{\mu}(z)\rangle=c\,\langle \phi(x) j^{\mu}(z)\rangle$. Borrowing the results in appendix \ref{Formulae}

\begin{equation}
 \langle \phi(x)\,j^{\mu}(z)\rangle=-i(d-2)C_d\frac{(x^{\mu}-z^{\mu})}{|x-z|^{d}}\qquad  \rightarrow \int d^dz \, \delta(z^0)\,\langle \phi(x)\,j^{0}(z)\rangle=-\frac{i}{2}{\rm sign}(x^0)\,.
\end{equation}

To $\mathcal{O}(\alpha)^2$ we need to compute $\langle e^{c\phi(x)}\,j^{\mu}(z_1)\,j^{\nu}(z_2)\rangle$. In this case (we only consider connected diagrams) 

\begin{eqnarray}
\langle e^{c\phi(x)}\,j^{\mu}(z_1)\,j^{\nu}(z_2)\rangle= \frac{c^2}{2}\langle \phi(x)^2 j^\mu (z_1) j^\nu (z_2)\rangle&=& c^2\,\langle \phi(x)\,j^{\mu}(z_1)\rangle\,\langle \phi(x)\,j^{\nu}(z_2)\rangle\,.
\end{eqnarray}
Since the $\mathcal{O}(\alpha^2)$ contribution involves an integral over $z_1,\,z_2$, it is clear that the two terms contribute the same and equal to the square of the $\mathcal{O}(\alpha)$ term. Thus, all in all to order $\mathcal{O}(\alpha^2)$ we find

\begin{equation}
\langle V(x)\,U_\alpha\rangle=1+\,\frac{c\,\alpha}{2}\,{\rm sign}(x^0)+\frac{1}{2}\,\Big(\frac{c\,\alpha}{2}{\rm sign}(x^0)\Big)^2+\cdots\sim e^{\,\frac{c\,\alpha}{2}\,{\rm sign}(x^0)}\,.
\end{equation}
Since in $d>2$ the shift symmetry is spontaneously broken, the 1-point functions change from side to side of the defect. Normalizing by the correlator for $x^0>0$ we find the expected action of the symmetry defect

\begin{equation}
\label{VU}
\frac{\langle V(x)\,U_\alpha\rangle}{\langle V(x)\,U_{\alpha}\rangle|_{x^0<0}}= e^{\,c\,\alpha\,\theta(x^0)}\,.
\end{equation}

We now turn to the more involved example of $\langle V(x)\,V(y)\,U_{\alpha}\rangle$. To order $\mathcal{O}(\alpha^0)$ this is simply $\dirac{V(x)\, V(y)}= e^{c^2\dirac{\phi(x)\phi(y)}}$.
To order $\mathcal{O}(\alpha)$ we need to compute $\langle e^{c\phi(x)}\,e^{c\phi(y)}\,j^{\mu}(z)\rangle$. Expanding the exponential one easily sees that

\begin{equation}
\langle V(x)\,V(y)\,j^{\mu}(z)\rangle=\left[c\,\langle \phi(x)\,j^{\mu}(z)\rangle+c\,\langle \phi(y)\,j^{\mu}(z)\rangle\right]\,e^{c^2\dirac{\phi(x)\phi(y)}} \,.
\end{equation}
Hence, borrowing previous results
\begin{equation}
\int d^dz\,\delta(z^0)\,\langle V(x)\,V(y)\,j^0(z)\rangle=-i\frac{c}{2}\big({\rm sign}(x^0)+{\rm sign}(y^0)\big)e^{c^2\dirac{\phi(x)\phi(y)}}\,.
\end{equation}
To order $\mathcal{O}(\alpha^2)$ we need to compute $\langle e^{c\phi(x)}\,e^{c\phi(y)}\,j^{\mu}(z_1)\,j^{\nu}(z_2)\rangle$. Upon expanding the exponentials\footnote{We omit terms with only $x$ or $y$ dependence, which will be cancelled by the denominators in eq. \eqref{VVU}.}
\begin{align*}
&\langle e^{c\phi(x)}\,e^{c\phi(y)}\,j^{\mu}(z_1)\,j^{\nu}(z_2)\rangle = \Big[\dirac{j^\mu(z_1)\, j^\nu(z_2)}+c^2 \langle \phi(x) j^{\mu}(z_1)\rangle\,\langle \phi(x) j^{\nu}(z_2)\rangle+\nonumber\\ 
&+c^2 \langle \phi(y) j^{\mu}(z_1)\rangle\,\langle \phi(y) j^{\nu}(z_2)\rangle+c^2 \langle \phi(x) j^{\mu}(z_1)\rangle\,\langle \phi(y) j^{\nu}(z_2)\rangle+c^2 \langle \phi(y) j^{\mu}(z_1)\rangle\,\langle \phi(x) j^{\nu}(z_2)\rangle \Big] e^{c^2\dirac{\phi(x)\phi(y)}}.
\end{align*}
Momentarily neglecting the contribution from $\langle j^{\mu}(z_1)\,j^{\nu}(z_2)\rangle$, we have that
\begin{equation}
\int d^dz_1\,\delta(z_1^0)\int d^dz_2\,\delta(z_2^0)\langle e^{c\phi(x)}\,e^{c\phi(y)}\,j^{0}(z_1)\,j^{0}(z_2)\rangle=-\Big(\frac{c}{2}\,\big({\rm sign}(x^0)+{\rm sign}(y^0)\big)\Big)^2\,.
\end{equation}
Then we would have the expected

\begin{equation}
\label{VVU}
%\frac{\langle V(x)\,V(y)\,U_{\alpha}\rangle}{\dirac{V(x)\,V(y)}}=e^{i \frac{c\alpha }{2}\,\big({\rm sign}(x^0)+{\rm sign}(y^0)\big)} \hspace{-0.7cm}\qquad \leadsto \qquad \hspace{-0.7cm}
\frac{\langle V(x)\,V(y)\,U_{\alpha}\rangle}{\langle V(x)\,U_{\alpha}\rangle|_{x^0<0}\,\langle V(y)\,U_{\alpha}\rangle|_{y^0<0}\dirac{V(x)\,V(y)}} =e^{c\,\alpha\,\big(\theta(x^0)+\theta(y^0)\big)} \,.
\end{equation}
To arrive to this result we have neglected the contribution to $\mathcal{O}(\alpha^2)$ of $\langle j^{\mu}(z_1)\,j^{\nu}(z_2)\rangle$. To compute this contribution note that the current-current correlator can be easily constructed taking derivatives of the 2-point function (see appendix \ref{Formulae})

\begin{equation}
\langle j^{0}(z_1)j^{0}(z_2)\rangle=-\frac{(d-2)C_d}{|z_1-z_2|^d}\,\Big(\delta^{00}-d\,\frac{(z_1-z_2)^{0}(z_1-z_2)^{0}}{|z_1-z_2|^2}\Big)\,.
\end{equation}
Then the neglected term is 
\begin{equation}
\,\int d^d z_1\, d^dz_2\,\delta(z_1^0)\,\delta(z_2^0)\,\langle j^{0}(z_1)j^{0}(z_2)\rangle=-\,\int d^{d-1} z_1 d^{d-1}z_2 \int \frac{d^d p}{(2\pi)^d} e^{ip\cdot (z_1-z_2)}\frac{p_0^2}{p^2}\,. 
\end{equation}
The integral over $z_1$ gives $\delta(\vec{p})$ leading to

\begin{equation}
-\int d^{d-1} z_2 \int \frac{dp^0}{2\pi}\,.
\end{equation}
We recognize in the (divergent) $p^0$ integral $\delta(0)$, so all in all we can write the neglected term as

\begin{equation}
- \int d^d z_2 \,\delta(z_2^0)^2 e^{c^2 \dirac{\phi(x)\phi(y)}}
\end{equation}
This would give an unexpected contribution to the $\langle V(x)\,V(y)\,U_{\alpha}\rangle$ correlation function. To remedy this, we can define a modified defect operator introducing $J_0=\alpha\,\delta(x^0)$ as

\begin{equation}
\widehat{U}_{\alpha}=U_{\alpha}\,e^{-\int d^dx \, \frac{1}{2}J_0^2}\,.
\end{equation}
The added term precisely cancels the offending contribution, leaving behind the result in eq. \eqref{VVU}. Note that the path integral with the insertion of $\widehat{U}_{\alpha}$ can be written as

\begin{equation}
\langle \mathcal{F}\,\widehat{U}_{\alpha}\rangle=\int \mathcal{D}\phi\,\mathcal{F}\,e^{-\int \frac{1}{2}D\phi^2}\,,\qquad D_{\mu}\phi=\partial\phi-J_{\mu}\,.
\end{equation}

\subsection{Global symmetries in scalar theories}\label{divergencescurrents}

We have seen that in both the shift and $U(1)$ symmetries, the naive definition of symmetry operators does not act as expected. Yet, upon adding a contact term, it is possible to find an improved version of the symmetry operator acting as expected on the charged operators. We now want to understand the origin of this contact term. Since it is clear that the relevant quantities are current-current correlators, let us construct a generating functional for them.

Focusing on the case of the complex scalar field with action

\begin{equation}
S=\int d^dx\, \partial\phi\partial\phi^\star+V(|\phi|)\,,
\end{equation}
the transformation $\phi\rightarrow e^{i\alpha}\phi$ is a symmetry. The infinitesimal transformation is

\begin{equation}
\delta\phi=\alpha\,i\phi\,,\qquad \delta\phi^\star=\alpha(-i\phi^\star)\,.
\end{equation}
As a consequence, the Noether conserved current is

\begin{equation}
j_0^{\mu}=\,\phi\partial^{\mu}\phi^\star-\phi^\star\partial^{\mu}\phi\,.
\end{equation}
When constructing a generating functional for current correlators, we would add a source for the current, which is effectively like adding a source term to the action

\begin{equation}
S\rightarrow S_J=S-i\int J_{\mu}\,j_0^{\mu}\,.
\end{equation}
However, since $j_0^{\mu}$ contains derivatives, the conserved current in $S_J$ is actually

\begin{equation}
j^{\mu}=j^{\mu}_0+2i\phi^\star\phi\,J^{\mu}\,.
\end{equation}
This motivates to construct instead

\begin{equation}
\widehat{S}_J=S_J+\int \phi^\star\phi\,J^{\mu}J^{\mu}\,.
\end{equation}
Clearly $\widehat{S}_J$ has the original $\phi\rightarrow e^{i\alpha}\phi$ symmetry for which the Noether current is $j^{\mu}$. Moreover, it holds that

\begin{equation}
\frac{1}{i}\frac{\delta \widehat{S}_J}{\delta J_{\mu}}=-j^{\mu}\,.
\end{equation}

Thus, the relevant generating functional for current correlators is

\begin{equation}
\widehat{Z}[J]=\int \mathcal{D}\phi\, e^{-\widehat{S}_J}\,,
\end{equation}
for which, by construction

\begin{equation}
\Big(\frac{1}{i}\frac{\delta}{\delta J_{\mu}(x)}\Big)\Big(\frac{1}{i}\frac{\delta}{\delta J_{\mu}(y)}\Big)\widehat{Z}[J] =\langle j^{\mu}(x)\,j^{\nu}(y)\rangle\,.
\end{equation}
Amusingly

\begin{equation}
\widehat{S}_J=\int D_{\mu}\phi(D^{\mu}\phi)^\star\,,\qquad D_{\mu}\phi=\partial_{\mu}\phi-i J_{\mu}\phi\,.
\end{equation}
The same holds for shift symmetries, where one finds

\begin{equation}
\widehat{Z}[J]=\int\mathcal{D}\phi\, e^{-\widehat{S}_J}\,,\qquad \widehat{S}_J=\int \frac{1}{2}D_{\mu}\phi D^{\mu}\phi\,,\qquad D_{\mu}\phi=\partial_{\mu}\phi-J_{\mu}\,.
\end{equation}

Thus to obtain the relevant functional to compute current-current correlators for conserved currents we need to add a background gauge field for the symmetry. Reading this in reverse, adding a background gauge field for the symmetry computes correlation functions for conserved currents. We can now see the relevance of this observation in our context. Consider for definiteness the case of a $U(1)$ symmetry (a similar story holds for the shift symmetry). The effect of the symmetry defect is to add a phase to the scalar as it crosses the defect: $\phi(x)\rightarrow e^{i\alpha\theta(x^0)}\phi(x)$. Hence, the effect of the defect is as like a (singular) background field for the global symmetry $J_{\mu}=\alpha\,\delta(x^0)$ \cite{Bah:2024ucp}. Thus, the correct definition for the defect is $\widehat{U}_{\alpha}$ rather that $U_{\alpha}$. 

Let us finally note that this discussion applies solely to bosons. For fermions with no derivative couplings, conserved currents do not involve derivatives and as a consequence the whole discussion above does not apply.

\section{Defect regularization by thickening and holography}\label{holography}

Very recently it has been proposed in \cite{Bergman:2024aly}  that defects implementing a continuous symmetry in holographic theories are realized holographically as non-BPS $D(q-1)$ branes living on the boundary. Indeed, in the the presence of $D(8-q)$ branes, the WZ action for non-BPS $D(q-1)$ branes is sensitive to the $D(8-q)$ charge and contains a hidden continuous parameter taking values in $U(1)$. As a result, after going to the tachyon vacuum, the non-BPS $D(q-1)$ brane leaves behind a phase proportional to the $D(8-q)$ brane charged linked times the hidden parameter, thus behaving as expected for a continuous symmetry operator. In turn, non-BPS $D(q-1)$-branes have a long history in String Theory. In particular, it is known that they can be realized through the decay of a $Dq/\overline{Dq}$ system, and it is then natural to ask whether this can be of interest in the context at hand. We have seen that the correct prescription to define symmetry defects for bosonic theories includes a contact term which can be easily traced back to the correct generating functional for (conserved) current correlators. We have regularized these contact terms through a cut-off as in eq. \eqref{regularizeddelta}, but of course one may use any other regularization, such as thickening the defect as in \cite{Bah:2024ucp}. It is natural to conjecture that this holographically corresponds to ``puffing up" the non-BPS $D(q-1)$ brane into a $Dq/\overline{Dq}$ system.  

\subsection{Kinks in $Dq/\overline{Dq}$ in $AdS_{d+1}$}

Let us investigate the $Dq/\overline{Dq}$ system in $AdS$. To that matter, we consider the general case of $AdS_{d+1}\times X_{9-d}$. In appropriate coordinates, the metric looks like

\begin{equation}
ds^2=R^2\,\left(\frac{dz^2+dx^2+ds_{\mathbb{R}^{1,d-1}}^2}{z^2}\right)+R^2\,ds_{X_{9-d}}^2\,.
\end{equation}
We now consider here a $Dq/\overline{Dq}$ acting as a codimension $p$ defect in the field theory directions. For the cases of interest $p<d-1$. As such, it wraps $\{\mathbb{R}^{1,d-p-2},\,z,\,C^{q-d+p+1}\}$, being $C^{q-d+p+1}$ a $q-d+p+1$-cycle in $X$. We assume that the only relevant fluctuations are those along the $x$ direction. Thus, the pull-back of the metric to each brane looks like ($\alpha=1,2$ stands for $Dq/\overline{Dq}$) 

\begin{equation}
ds^2_{\alpha}=\frac{R^2}{z^2}\,d\xi^I\,d\xi^J\,\mathcal{G}^{\alpha}_{IJ}+R^2\,ds_{C^{q-d+p+1}}^2\,,
\end{equation}
where the $I$ coordinate can take the values $(z,i)$ ($i=1,...,d-p-1$) and
\begin{equation}
d\xi^I\,\mathcal{G}^{\alpha}_{IJ}\,d\xi^J=dx^idx^j\,(\eta_{ij}+\partial_ix_{\alpha}\partial_jx_{\alpha})+dz^2\,(1+\partial_zx_{\alpha}^2)+2\,dz\,dx^i\,\partial_zx_{\alpha}\partial_ix_{\alpha}\,.
\end{equation}

The DBI action for the $Dq/\overline{Dq}$ is then \cite{Sen:2003tm}

\begin{equation}
S=-\int V(\mathcal{T}, x_1-x_2)\,(\sqrt{-{\rm det}\mathbf{A}_1}+\sqrt{-{\rm det}\mathbf{A}_2})\,,
\end{equation}
being

\begin{equation}
(\mathbf{A}_{\alpha})_{IJ}=\frac{R^2}{z^2}\,\mathcal{G}^{\alpha}_{IJ}+F^{\alpha}_{IJ}+\frac{1}{2}\overline{D_I\mathcal{T}}D_J\mathcal{T}+\frac{1}{2}\overline{D_J\mathcal{T}}D_I\mathcal{T}\,,
\end{equation}
 and

\begin{equation}
D_I\mathcal{T}=\partial_I\mathcal{T}-i (A^1_I-A^2_I) \mathcal{T}\,.
\end{equation}
The form of the potential $V=V(|\mathcal{T}|,\,x_1-x_2)$ suggests to consider the fields

\begin{equation}
X=\frac{x_1+x_2}{2}\,,\qquad Y=\frac{x_1-x_2}{2}\,.
\end{equation}
The field $X$ only appears through derivatives, and has no potential, and can be interpreted as the Goldstone mode associated to the translational symmetry broken by the system itself. In turn, for small $\mathcal{T}$ the rough form of the tachyon potential is

\begin{equation}
V\sim 1+|\mathcal{T}|^2\Big( \frac{Y^2}{4}-1\Big)+\cdots\,.
\end{equation}
Since there is a tachyonic mass term for $\mathcal{T}$, it is natural to expect that $\mathcal{T}$ will roll down its potential, thus giving effectively a mass to $Y$ through the term $\frac{1}{4}Y^2 |\mathcal{T}|^2$. As a consequence, $Y$ will be effectively set to zero, which sets $x_1=x_2=x$. Let us in addition assume the defect to be at a fixed transverse position and set to zero the gauge fields. Moreover, consider the real tachyon profile $\mathcal{T}=T$ \cite{Sen:1999mg}, with $T=T(z)$ in order not to break the boundary Poincaré symmetry. Then, the action boils down to

\begin{equation}
S=-2\,{\rm Vol}(C^{q-d+p+1})R^{q+1}\int dz\frac{V(T)}{z^{d-p}}\,\sqrt{1+\frac{z^2}{R^2}\,\partial_zT^2}\,.
\end{equation}
Introducing $x=\log \frac{z}{R}$. The action becomes

\begin{equation}
S=-2\,{\rm Vol}(C^{q-d+p+1})R^{q-d+p+2}\int dx\,e^{-(d-p-1)x}\,V(T)\,\sqrt{1+R^{-2}\,\partial_xT^2}\,.
\end{equation}
The equation of motion is

\begin{equation}\label{TachyonEOMs}
\partial_x\Big(\frac{V\,\partial_xT}{R^2\,\sqrt{1+R^{-2}\partial_xT^2}}\Big)-(d-p-1)\,\frac{V\,\partial_xT}{R^2\,\sqrt{1+R^{-2}\partial_xT^2}}-\frac{\partial V}{\partial T}\,\sqrt{1+R^{-2}\partial_xT^2}=0\,.
\end{equation}
Note first that a solution to this equation is $T\rightarrow \infty$, since $V=0$. This corresponds to the branes coinciding --since $Y=0$ as argued above-- $Dq/\overline{Dq}$ which annihilate. In order to investigate whether other solutions exist let us suppose that there is a region where the tachyon and its derivatives are small. Assuming, for definiteness, popular forms for the tachyon potential \textit{e.g.} \cite{Kutasov:2003er,Bergman:2007pm} 

\begin{equation}
V=\frac{1}{\cosh(c_1T)}\,, \qquad{\rm or} \qquad V=e^{-c_2^2\,T^2}\,,
\end{equation}
($c_{1,2}$ are dimensionful constants, proportional to $\ell_s$, whose precise value is immaterial for our purposes) one easily finds that 

\begin{equation}
T=T_0\, e^{(d-p-1)\,x}\,F(x)
\end{equation}
where $F(x)$ is a bounded function of $x$. We see that for large negative $x$, which corresponds to the region close to the $AdS$ boundary, indeed the tachyon and its derivatives are exponentially small since $p<d-1$. Thus, in that region, the tachyon is at the top of its potential and the $Dq/\overline{Dq}$ have not annihilated. However, as one goes deep into the bulk, the tachyon grows, departing the range of validity of the approximation. In that region the only solution is the $T\rightarrow \infty$ discussed above. Hence, as the $Dq/\overline{Dq}$ go into the bulk, the tachyon goes to its minimum, where the $Dq/\overline{Dq}$ have annihilated. This picture is also consistent with the behavior of the energy-momentum tensor. Its non-zero components are

\begin{equation}\label{EMtensor}
T^{zz}=-\frac{V(T)}{\sqrt{1+\frac{z^2}{R^2} (\partial_zT)^2}}\frac{z^2}{R^2} \,,\qquad T^{ij}=\left(1+\frac{z^2}{R^2}(\partial_z T)^2\right)T^{zz}\eta^{ij}\,.
\end{equation} 
In $AdS$, the energy-momentum tensor is covariantly conserved $\nabla_\mu T^{\mu\nu}=0$. Upon writing the equation in terms of the logaritmic coordinate $x=\log\frac{z}{R}$, this reduces to

\begin{equation}
\partial_x \log T^{zz}=2-(d-p-1)R^{-2}(\partial_x T)^2\,.
\end{equation}

We can then study the equation in the regime of small derivatives, which reduces to 

\begin{equation}
\partial_x \log T^{zz}= 2\,\,\,\rightarrow\,\,\, T^{zz}\sim C\frac{z^2}{R^2}\,.
\end{equation}
This is compatible with the form of the energy-momentum tensor in eq. \eqref{EMtensor} in the limit in which $V(T)\sim 1$ and $R^{-1}\partial_x T\ll 1$.  

In turn, in the large derivative limit $R^{-1}\partial_x T\gg \sqrt{\frac{2}{d-p-1}}$, we can recast the conservation equation in terms of an integral equation 

\begin{equation}
T^{zz}\sim C'e^{- (d-p-1)\int^z  \frac{z^2}{R^2}(\partial_z T)^2 dz}\,.
\end{equation}
In this regime, the energy momentum tensor is exponentially suppressed in terms of the integral of the square derivative of the tachyon. This is compatible with the system being at the minimum of the potential in the large derivative region.

Let us come back to the dynamics of $Y$. We argued that as the tachyon rolls down, $Y$ gets a mass and gets frozen to 0. However, there is some region close to the boundary where the tachyon field is small, so $Y$ can not be set to zero and the $Dq/\overline{Dq}$ can be separated. Since this happens close to the boundary where the tachyon is small, we can qualitatively study this region by setting to zero the tachyon and consider only the scalar fluctuations.  Assuming only $z$-dependence, the action for $Y$ reduces to the action of a $D-$brane hanging towards the $AdS$ interior

\begin{equation}
\label{SUshaped}
S=-2\,V(0){\rm Vol}(C^{q-d+p+1})\,R^{q+1}\,\int dz\, \frac{1}{z^{d-p}}\,\sqrt{1+\partial Y^2}\,.
\end{equation}
The EOMs are then

\begin{equation}
\partial_z\left( \frac{\partial_zY}{z^{d-p}\,\sqrt{1+\partial Y^2}}\right)=0\,.
\end{equation}

For small $z$ the solution is $Y\sim Y_0$ (see appendix \ref{appU}). This admits a heuristic explanation: due to the $AdS$ warp factor, kinetic terms in $AdS$ are strongly suppressed, and as a consequence fields become effectively non-dynamical close to the boundary. Note that the same argument applies to the overall translational mode: its kinetic term is suppressed and on the boundary $X$ is also fixed.

Thus, putting all the pieces together, close to boundary, the $Dq/\overline{Dq}$ is at fixed constant separation. In turn, as the branes penetrate in $AdS$, they approach, the tachyon grows and eventually they coincide and annihilate. As a consequence, we can regard the $Dq/\overline{Dq}$ system as recombinating into a single $U$-shaped $q$-brane hanging from the boundary. This system can be exactly solved, as we review in appendix \ref{appU}. Moreover, in the regime where the $U$ is small --which corresponds to a very little separation in the $Y$ direction, we expect the kinetic terms for worldvolume fluctuations to be suppressed. It is then natural to identify this $U$-shaped $Dq$, in the regime where the $U$ is pushed to the boundary (that is, the $U$ hangs very little from the boundary),  with the non-BPS $D(q-1)$ brane realizing the symmetry defect.

\subsection{The baryonic symmetry in Klebanov-Witten}

In order to support the conjectured identification of the (small) $U$-shaped $Dq$ with a continuous symmetry operator, let us focus on the specific example of the baryonic symmetry in the Klebanov-Witten theory discussed in \cite{Bergman:2024aly}. 

The Klebanov-Witten theory is a 4d ${\cal N}=1$ supersymmetric gauge theory with gauge group $SU(N)\times SU(N)$ and chiral multiplets $A_i, \,\, B_i$ in the $({\bf N},\bar{\bf N})$ and $(\bar{\bf N},{\bf N})$ representations respectively \cite{Klebanov:1998hh}. The theory has a baryonic $U(1)$ global symmetry under which the fields $A_i$ and $B_i$ carry opposite charge. While the usual mesons are neutral under this symmetry, one can construct determinant-like gauge-invariant operators which, in the appropriate normalization, carry unit charge under the baryonic symmetry \cite{Gubser:1998fp}. The Klebanov-Witten theory admits a holographic dual in terms of Type IIB string theory on $AdS_5 \times T^{1,1}$ with $N$ units of $\tilde{F}_5$ flux. Topologically, $T^{1,1}\sim S^2\times S^3$. Then, the baryonic $U(1)$ symmetry is dual to a $U(1)$ gauge field in $AdS_5$ that comes from reducing the RR 4-form potential $C_4$ on the $S^3$ as $C_4\sim A_B\wedge \omega_{S^3}$, so that 

\begin{equation}
\label{AB}
dA_B=\int_{S^3}\tilde{F}_5\,.
\end{equation}

In turn, the baryon operators are dual to a $D3$-brane wrapping $S^3$, extending in the radial direction, and ending on the boundary of $AdS_5$. In \cite{Bergman:2024aly} it was proposed that the symmetry operator associated to the baryonic symmetry corresponds to a non-BPS $D4$-brane that wraps the $S^2$,
extends along a 3-manifold $M_3$ in spacetime, and is taken to the boundary of $AdS_5$. This $D4$ brane naturally links with the baryonic $D3$ brane and carries a hidden parameter on its worldvolume leaving behind the expected phase for the action of a continuous symmetry operator.

\begin{figure}[h!]
\centering
\includegraphics[scale=.18, trim={0.2cm, 9cm, 0.5cm, 9cm}, clip]{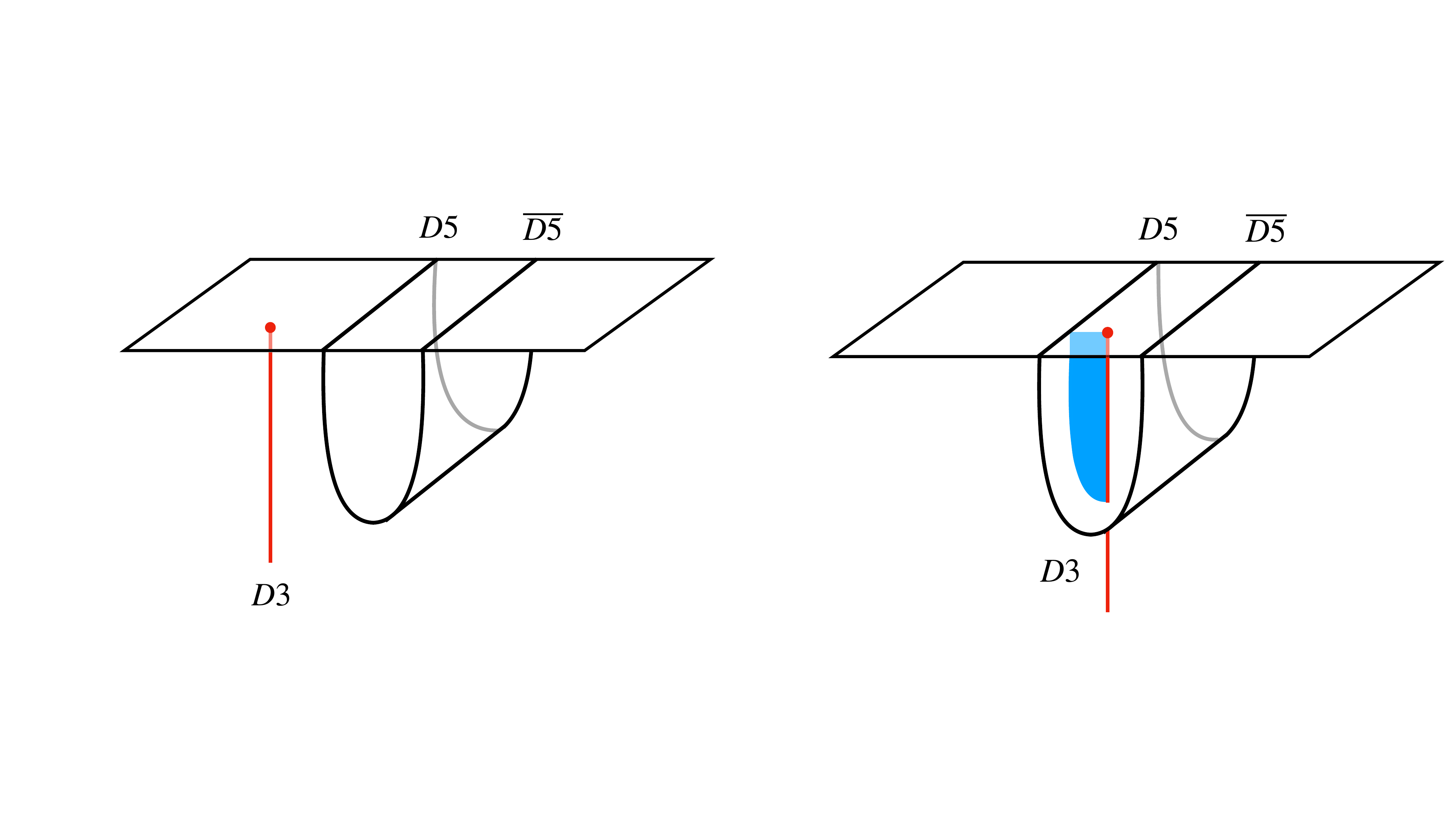}
\caption{A string is created when a $D3$ brane crosses the $D5/\overline{D5}$ system.}
\label{stringcreation}
\end{figure}

As proposed in the previous section, we may regard the (fattened) non-BPS $D4$ as a $D5/\overline{D5}$ system. We now want to argue that this offers further insight into the proposed $D4$ as a symmetry defect. To that matter, we first consider a single $D5$ brane in $AdS_5\times T^{1,1}$ wrapping $M_3$ in the boundary, $z$ and the $S^2$ in the $T^{1,1}$. Such brane produces a RR 3-form field strength flux with one unit over the $S^3$. As a consequence, as a baryonic $D3$ is dragged across the $D5$ it develops a worldvolume tadpole which must be cancelled by attaching one fundamental string. This allows to interpret the $D5$ as a domain wall from $SU(N)\times SU(N)$ to $SU(N)\times SU(N+1)$ \cite{Gubser:1998fp}. The addition of the extra $\overline{D5}$ brane brings us back to $SU(N)\times SU(N)$, so that the combined system $D5/\overline{D5}$ system can be regarded as a defect which is only sensitive to the baryons: as baryonic $D3$ branes are dragged across, a F1 is created filling the resulting $U$-shaped $D5$, see fig. \ref{stringcreation}.

 Moreover, since the $U$ shaped D5 is assumed to be small, kinetic terms for worldvolume fluctuations are frozen and the system becomes topological.

We can make explicit the relation to the baryonic symmetry. Let us consider first the case of a single $D5$ brane. Such brane involves a worldvolume coupling of the form 

\begin{equation}
S_{D5}\supset -T_5\int C_4\wedge da\,,
\end{equation}
where $a$ is the worldvolume gauge field on the $D5$ brane. As a consequence, in the presence of the $D5$, the equation of motion for $C_4$ fluctuations is of the form

\begin{equation}
d F_5 = da\wedge \delta_{D5}\,,
\end{equation}
where $\delta_{D5}$ is a 4d Dirac delta supported on the worldvolume of the $D5$ brane $M_3$, $z$ and $S^2$. Thus, $\delta_{D5}$ has components along the transverse direction $x$ to the defect (localized at $x=0$)  and the $S^3$. This can be written as $d(F_5-a\wedge \delta_{D5})=0$. This shows that $C_4$ must transform under $a$ gauge transformations. Integrating $F_5-a\wedge \delta_{D5}$ on the $S_3$ we find (omitting unimportant numerical factors)

\begin{equation}
\int_{S^3}F_5-a\wedge \delta_{D5}=\int_{S^3}F_5-\delta(x)\,a\wedge dx\,.
\end{equation}
Therefore, in view of eq. \eqref{AB}, the worldvolume gauge transformation $\delta a=d\lambda$ induces a shift gauge transformation for the baryonic field as $\delta A_B=\delta(x) \lambda dx$. In modern parlance, the worldvolume gauge field on the $D5$ represents the gauge field for the shift symmetry of $A_B$, which is the holographic realization of the $U(1)$ baryonic symmetry. In our context, as the baryon $D3$ crosses the $D5/\overline{D5}$, a string is created. This in particular excites a worldvolume gauge field on the $D5/\overline{D5}$ corresponding to the endpoints of the string --which look like worldvolume electron/positron. As the branes are brought together (corresponding to the very small $U$ limit), these fields cancel each other up to an arbitrary gauge transformation. Through the anomaly this gauge transformation becomes physical and generates a $\delta A_B\sim d(\lambda\,\theta(x))$, reproducing the field theory discussion.

\section{Conclusions}\label{conclusions}

In this work we have studied continuous symmetry operators. In the first part we have explicitly computed correlation functions with defect insertions in (mostly free) QFT. The upshot is that the naive definition of the defect must be supplemented with contact terms in order to reproduce the correct action of the symmetry defect. This probably comes at no surprise, since, as we argued, it can be traced back to the fact that the correct generating functional for current correlators is identical to coupling the QFT to a background gauge field. While this is clear for fermions, for bosons, which have kinetic terms quadratic in derivatives,  it comes with an extra contribution whose effect is to precisely produce the required contact terms. 

In the second part of this work, we have proposed that holographic symmetry defects, described by non-BPS $D(q-1)$ branes, may be regarded as $Dq/\overline{Dq}$ systems in the limit of small separation (when the worldvolume fluctuations are frozen). Focusing on the case of the baryonic symmetry in Klebanov-Witten, it is possible to explicitly see how the symmetry operator is sensitive to the baryon symmetry. Indeed, regarding the $D4$ as $D5/\overline{D5}$, as baryons are dragged across the defect a string is created and left behind filling the $U$-shaped configuration. Moreover, by a similar argument to \cite{Green:1996dd}, the baryonic symmetry is mapped to the worldvolume gauge field on the $D5/\overline{D5}$ system, in parallel with the field theory observation that the effect of the defect is akin to a background gauge field for the symmetry.

A natural question stemming from our work concerns the generality of $U$-shaped $D$-branes. These often appear in a number of holographic realizations of QFT. It is natural to ask whether, at least in the limit of small $U$, these can be regarded as symmetry operators. A prime example are massive flavors in Witten-Sakai-Sugimoto \cite{Witten:1998zw, Sakai:2004cn}, which are realized in terms of a $D8/\overline{D8}$ system recombining into a $U$-shaped $D8$ hanging from the cigar geometry \cite{Bergman:2007pm}. Clearly, if a $D0$ brane wrapping the $S^1$ of the cigar is pushed to the boundary across the $U$, a fundamental string is created. The same string was considered in \cite{Aharony:2008an}, where it was argued that it captures the quark mass, and thus can be regarded as a shift of the $\theta$ angle. In turn, the $D0$ brane is identified with a field theory instanton and the would-be non-BPS $D7$ resulting from the $D8/\overline{D8}$ would realize the $-1$ form symmetry shifting the $\theta$ angle just as in the $\mathcal{N}=4$ SYM case described in \cite{Bergman:2024aly}.\footnote{Note that the dilaton and metric factors consipire so as to keep the tension of the $D0$ finite even at the boundary.} It would be very interesting to further study this --and similar systems-- to explore the mechanism proposed in this work.

\section*{Acknowledgments}

We would like to thank O. Bergman and E. Garc\'ia Valdecasas for early collaborations and discussions, as well as E.Kiritsis for useful discussions. D.R-G. is grateful to the Physics Department of the University of California at San Diego, the CMSA at Harvard University as well as to the organizers of the program on \textit{Symmetries and Gravity} for hospitality. The authors are supported in part by the Spanish national grant MCIU-22-PID2021-123021NB-I00.

\begin{appendix}

\section{Useful formul\ae}\label{Formulae}

For Fourier-transforms we use

\begin{equation}
\frac{1}{(x^2)^{\alpha}}=\frac{(4\pi)^{\frac{d}{2}}\,\Gamma(\frac{d}{2}-\alpha)}{4^{\alpha}\,\Gamma(\alpha)}\,\int \frac{d^dp}{(2\pi)^d}\,\frac{e^{ipx}}{(p^2)^{\frac{d}{2}-\alpha}}\,.
\end{equation}
In particular, the propagator in $\mathbb{R}^d$ for a complex scalar theory reads

\begin{equation}
\dirac{\phi(x)\phi^\star(0)}=\int\frac{d^dp}{(2\pi)^d}\, \frac{e^{ipx}}{p^2}=\frac{C_d}{|x|^{d-2}}\,,\qquad C_d=\frac{1}{4\pi^{\frac{d}{2}}}\,\Gamma\left(\frac{d}{2}-1\right)\,.
\end{equation}
Note that we may integrate over $p^0$ to obtain

\begin{equation}
\frac{C_d}{|x|^{d-2}}=\int\frac{d^{d-1}\vec{p}}{(2\pi)^{d-1}}\, \frac{e^{-|\vec{p}|\,|x^0|+i\vec{p}\cdot\vec{x}}}{2|\vec{p}|}\,.
\end{equation}
Moreover, taking derivatives of the propagator one finds

\begin{eqnarray}
&&\partial^\mu \dirac{\phi(x)\phi^\star(0)}= \int\frac{d^dp}{(2\pi)^d}\, \frac{i p^{\mu} e^{ipx}}{p^2}=-\frac{\,(d-2)\,C_d}{|x|^d}\,x_{\mu}\,; \\ 
&& \partial^\mu \partial^\nu \dirac{\phi(x)\phi^\star(0)} =-\int\frac{d^dp}{(2\pi)^d}\, \frac{p^{\mu} p^{\nu} e^{ipx}}{p^2}=-\frac{(d-2)C_d}{|x|^d}\,	\left(\delta^{\mu\nu}-d\,\frac{x^{\mu}x^{\nu}}{|x|^2}\right  )\,.
\end{eqnarray}

\section{Calculations of integrals}\label{Calculationsofintegrals}

We report here the calculations of the integrals in eq. \eqref{Integral1}. The integral

\begin{equation}
(d-2)C_d^2\int d^{d}z\delta(z_0) \left(\frac{x^0}{|x-z|^d|y-z|^{d-2}}-\frac{y^0}{|x-z|^{d-2}|y-z|^d}\right)
\end{equation}
can be calculated transforming in momentum space and back. In particular, the first integral in momentum space reads

\begin{equation}
-i\int d^d z\delta(z_0) \int \frac{d^dp}{(2\pi)^d} \frac{d^d q}{(2\pi)^d} e^{i p\cdot (x-z)}e^{iq\cdot (y-z)} \frac{p_0}{p^2q^2}\,.
\end{equation}
Shifting $\vec{z}\rightarrow \vec{z}+\vec{x}$ and integrating over $z$, we get 

\begin{equation}
\int d^{d-1} z e^{-i(p+q)\cdot z}= \delta(\vec{p}+\vec{q})\,.
\end{equation}

We then obtain

\begin{equation}
-i\int \frac{d^dq}{(2\pi)^d} \frac{dp_0}{2\pi} e^{i\vec{q}\cdot (\vec{y}-\vec{x})}e^{ip_0x^0+iq_0y^0}\frac{p_0}{(p_0^2+\vec{q}^2)(q_0^2+\vec{q}^2)}\,.
\end{equation}
Performing the integral over $p_0$ reads 

\begin{equation}
-\frac{e^{i\vec{q}\cdot (\vec{y}-\vec{x})}}{q^2}e^{iq_0 y^0}\partial_0^x \left(\frac{e^{-|\vec{q}| |x^0|}}{2|\vec{q}|}\right)=\frac{1}{2q^2}e^{i\vec{q}\cdot (\vec{y}-\vec{x})} e^{i q_0 y^0-|\vec{q}||x^0|} \text{sgn}(x^0)\,.
\end{equation}
Integrating also over $q_0$, we obtain
\begin{equation}
\frac{1}{2}\text{sgn}(x^0)\int \frac{d^{d-1}q}{(2\pi)^{d-1}}e^{i \vec{q}\cdot (\vec{y}-\vec{x})}e^{-|\vec{q}||x^0|-|\vec{q}||y^0|}\frac{1}{2|\vec{q}|}\,.
\end{equation}
The second integral can be calculated as well by just exchanging $x\leftrightarrow y$

\begin{equation}
-\frac{1}{2}\text{sgn}(y^0)\int \frac{d^{d-1}q}{(2\pi)^{d-1}}e^{i \vec{q}\cdot (\vec{y}-\vec{x})}e^{-|\vec{q}||x^0|-|\vec{q}||y^0|}\frac{1}{2|\vec{q}|}\,.
\end{equation}

Using results from appendix \ref{Formulae}, the sum of the integrals reads

\begin{equation}
 \frac{1}{2}(\text{sgn}(x^0)-\text{sgn}(y^0))\int \frac{d^{d-1}q}{(2\pi)^{d-1}}e^{i \vec{q}\cdot (\vec{y}-\vec{x})}e^{-|\vec{q}||x^0|-|\vec{q}||y^0|}\frac{1}{2|\vec{q}|}=\dirac{\phi(x)\phi^\ast(y)} \frac{1}{2}(\text{sgn}(x^0)-\text{sgn}(y^0))\,.
\end{equation}

\section{$U$-shaped $p$-branes in $AdS$}\label{appU}

In this section we review $U$-shaped $p$-branes in $AdS$. Consider $AdS_{d+1}$ with metric

\begin{equation}
ds^2=\frac{d\vec{x}_{p}^2+d\vec{x}_{d-p-1}^2+dY^2+dz^2}{z^2}\,.
\end{equation}
We wrap a brane on $\{\mathbb{R}^{d-p-1},\,z\}$ and assume $Y=Y(z)$, fixing the boundary condition that one enpoint of the system is at $Y(0)=- \frac{L}{2}$, while the other is at $Y(0)=+\frac{L}{2}$. The action coincides with eq. \eqref{SUshaped}. Explicitly

\begin{equation}
S=-T\int d^{p+1}\vec{x}\,dz\,\frac{1}{z^{d-p}}\,\sqrt{1+Y'^2}\,.
\end{equation}
The equation of motion can be reduced to

\begin{equation}
Y'=\pm \frac{c z^{d-p}}{\sqrt{1-c^2\,z^{2d-2p}}}\,,\qquad c=\frac{1}{z_{\rm max}^{d-p}}
\end{equation}
with $z_{\rm max}$ a constant. Clearly $z\leq z_{\rm max}$, so $z_{\rm max}$ denotes the maximal depth in $AdS$ attained by the brane. The solution to the equation of motion is

\begin{equation}
\label{Yeom}
Y=\pm z_{\rm max}\,\int_{0}^{\frac{z}{z_{\rm max}}}\,du\, \frac{u^{d-p}}{\sqrt{1-u^{2d-2p}}}\,.
\end{equation}
This can be integrated 

\begin{equation}
Y=\pm z_{\text{max}}\frac{\left(\frac{z}{z_\text{max}}\right)^{d-p+1} \, _2F_1\left(\frac{1}{2},\frac{d-p+1}{2 d-2p};\frac{d-p+1}{2 d-2p}+1;\left(\frac{z}{z_\text{max}}\right)^{2 d-2p}\right)}{d-p+1}\,.
%\mp\frac{\sqrt{\pi } \Gamma \left(\frac{3 d-3p-3}{2 d-2p}+\frac{4}{2 d-2p}\right)}{(d-p+1) \Gamma \left(\frac{2 d-2p}{2 d-2p}+\frac{3}{2 d-2p}\right)}\,.
\end{equation} 
By symmetry, the turning point --and the point where the maximal depth in $AdS$ is attained-- is at $Y=0$. Hence, choosing the positive branch, from eq. \eqref{Yeom}

\begin{equation}
\frac{L}{2}=z_{\rm max} \int_{0}^{1}\,du\, \frac{u^{d-p}}{\sqrt{1-u^{2d-2p}}}=z_{\rm max}\frac{\sqrt{\pi}\, \Gamma\left(\frac{d-p+1}{2(d-p)}\right)}{\Gamma\left(\frac{1}{2d-2p}\right)}\,.
%z_{\rm max} \frac{\sqrt{\pi}\,\Gamma(\frac{1+3d-3p}{2d-2p})}{(d-p+1)\,\Gamma(1+\frac{1}{2d-2p})}\,.
\end{equation}
Note in particular that as $L\sim z_{\rm max}$, so very little separated $U$'s deep very little into $AdS$.

\end{appendix}

\bibliography{ArXiv_v2bib}
\bibliographystyle{JHEP}

\end{document}